%% file: main.tex
\documentclass[11pt,a4paper]{article}
\usepackage{times,latexsym}
\usepackage{url}
\usepackage[T1]{fontenc}

\usepackage[acceptedWithA]{tacl2018v2}

\usepackage{xspace,mfirstuc,tabulary}
\usepackage{graphicx}
\usepackage{multirow}
\usepackage{multicol}
\usepackage{lipsum}
\usepackage{subcaption}
\usepackage{subfloat}
\usepackage{amsmath}
\usepackage{caption}
\usepackage{color}
\usepackage{xcolor}
\usepackage{hyperref}

\newcommand\Tlide{\char`\~}

\newif\iftaclinstructions
\taclinstructionsfalse 
\iftaclinstructions
\renewcommand{\confidential}{}
\renewcommand{\anonsubtext}{(No author info supplied here, for consistency with
TACL-submission anonymization requirements)}
\newcommand{\instr}
\fi

\iftaclpubformat 

\else

\fi

\title{\textit{The Word2vec Graph} Model for Author Attribution and Genre Detection in Literary Analysis}

\author{Nafis Irtiza Tripto\textsuperscript{1,2},  Mohammed Eunus Ali\textsuperscript{2}\\
        \textsuperscript{1}The Pennsylvania State University, USA\\ 
       \textsuperscript{1}\texttt {nit5154@psu.edu} \\
         \textsuperscript{2}Bangladesh University of Engineering and Technology, Bangladesh   \\ 
          \textsuperscript{2}\texttt {eunus@cse.buet.ac.bd} 
         }

\date{}

\begin{document}
\maketitle
\begin{abstract}
Analyzing the writing styles of authors and articles is a key to supporting various literary analyses such as author attribution and genre detection. Over the years, rich sets of features that include stylometry, bag-of-words, n-grams have been widely used to perform such  analysis. However, the effectiveness of these features largely depends on the linguistic aspects of a particular language and datasets specific characteristics. Consequently, techniques based on these feature sets cannot give desired results across domains. In this paper, we propose a novel \emph{Word2vec graph} based modeling of a document that can rightly capture both context and style of the document. By using these \textit{Word2vec graph} based features, we perform classification to perform author attribution and genre detection tasks. Our detailed experimental study with a comprehensive set of literary writings shows the effectiveness of this method over traditional feature based approaches. Our code and data are publicly available\footnote{\url{https://cutt.ly/svLjSgk}}.
\end{abstract}

\input{sec1_introduction}

\input{sec2_related_works}

\input{sec4_Methodology}
\input{sec5_experiment}

\input{sec6_conclusion}

\bibliography{main}
\bibliographystyle{acl_natbib}

\end{document}

%% file: sec1_introduction.tex
\section{Introduction}
\label{sec:intro}

Research on writing styles, which is commonly referred as stylometry, has received considerable attention from both academia and industry over the years due to its impeccable impact on various practical applications in natural language processing (NLP) and literary analysis. Example applications include author attribution: identifying the author of a particular text~\cite{kevselj2003n,zhao2007searching}, genre detection: assigning the correct literary genre such as novels and short stories to a document based on its text and other attributes~\cite{kessler1997automatic}, stylochronometry: writing style change of an author over time~\cite{can2004change}, author verification: verifying whether two documents are written by the same author~\cite{halvani2016authorship}. All these tasks can be generally considered as a text classification problem that assigns a label, i.e., author/genre/chronological timeline, to a text of various forms such as story, article, or writing\footnote{Throughout the paper, we use the term \textit{document} to represent
any literary writing, story or article to avoid ambiguity.}

\begin{figure*}[ht]
	\centering
	\begin{subfigure}{0.66\columnwidth}
		\includegraphics[scale=0.3]{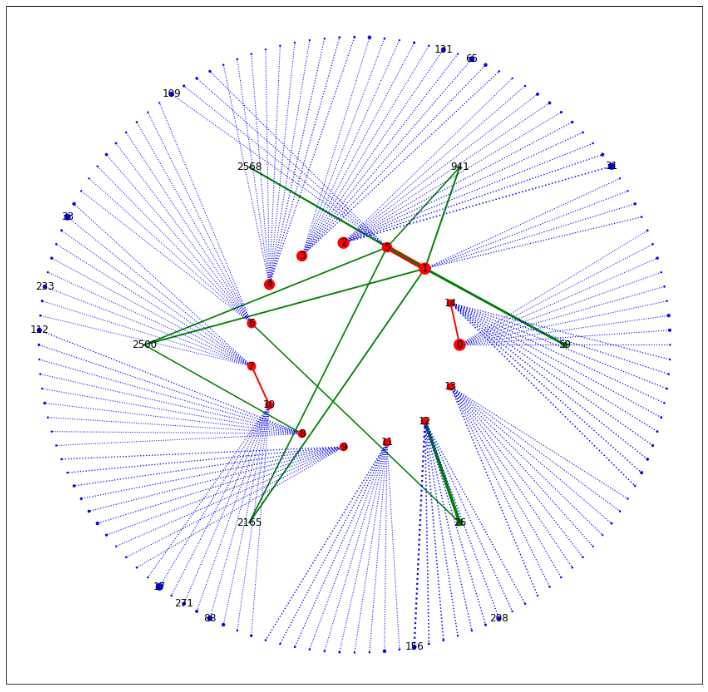}
		\caption{Historical novel
		(\textit{Modhannyo})}
		\label{subfig-long}
	\end{subfigure}%
	\begin{subfigure}{0.66\columnwidth}
		\includegraphics[scale=0.3]{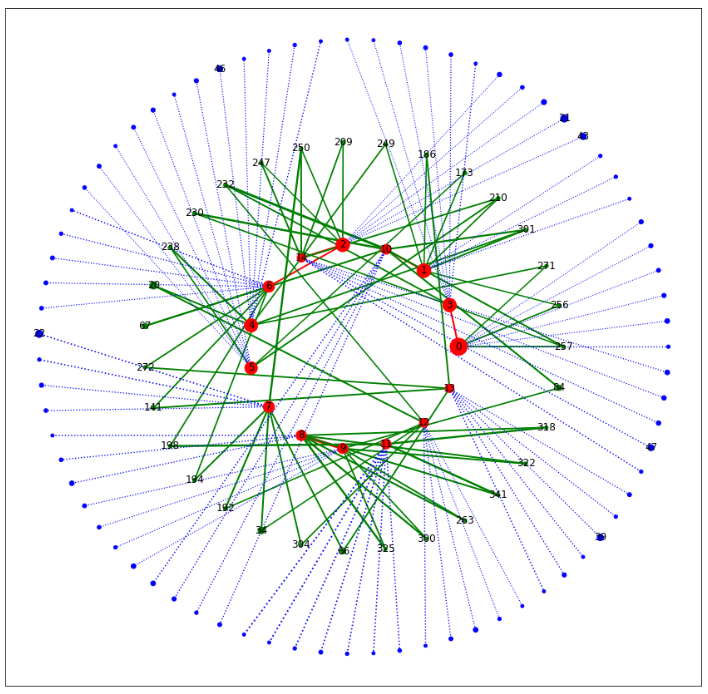}
		\caption{Novel (\textit{Nondito Noroke})}
		\label{subfig-novel}
	\end{subfigure}%
	\begin{subfigure}{0.66\columnwidth}
		\includegraphics[scale=0.3]{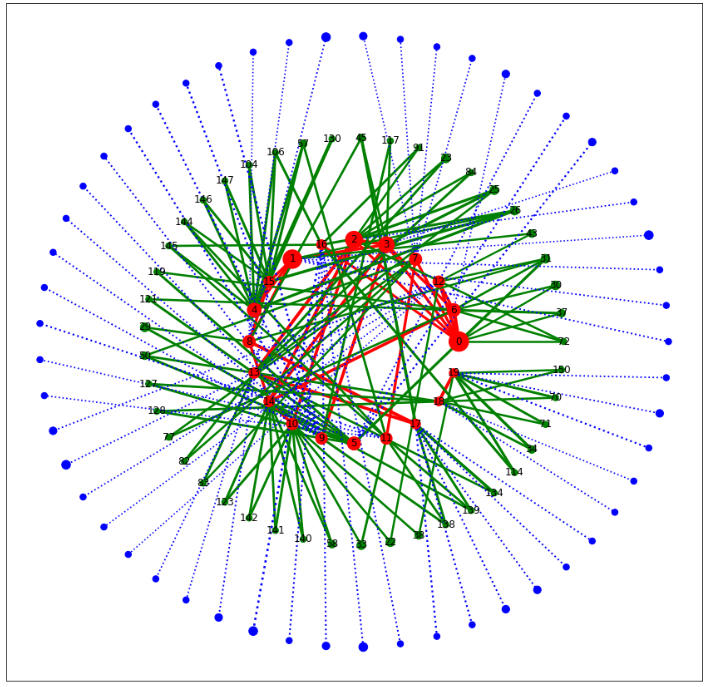}
		\caption{Short story (\textit{Elebele})}
		\label{subfig-short}
	\end{subfigure}%
	\caption{\textit{Word2vec graph} on three writings of Humayun Ahmed. 
	The red, green and black color represent \textit{core, multi, boundary} nodes/edges respectively (discussed  in  Section~\ref{sec-method_w2v}). The \textit{core} and \textit{multi} nodes are densely connected for short story. There are limited \textit{core} edges in both novel and  historical novel
	}
	\label{fig-example}
\end{figure*}

 To  classify a text in the literature and other related domains, a wide range of feature sets, such as bag-of-words, word n-grams, lexical, syntactic, semantic, structural attributes, and sequential modeling have been used~\cite{neal2018surveying}.  However, utilizing these distinct sets of features for classification has some key shortcomings, in particular for literary analysis of a specific language. These are: (i) A literary writing itself is not well structured like Wikipedia/newspaper article or online reviews. Instead of providing concise and straightforward information, a literary writing conveys intricate themes throughout a very long narrative~\cite{worsham2018genre}. 
 Therefore, structural features, such as paragraph length or indentations/URLs that rely on the formation of the document are inappropriate for this domain.
 (ii)  Many syntactic, semantic features, \textcolor{black}{such as synonyms or semantic dependencies,}  are language-specific, and extracting them from text require specific procedures and understanding for that particular language.
Also, the efficiency of these features often depends on the characteristics of the datasets~\cite{sari2018topic}.
Thus a feature that works for a specific language may not be appropriate for the literature domain of  other languages. 

To overcome these limitations in the literary analysis, we present a novel word-embedding graph (namely \textit{Word2vec graph}) that can capture the underlying representation of a document. Each node of the \textit{Word2vec graph} represents a word in the document and an weighted edge between two words denote their similarities in the embedding space. 
\textcolor{black}{
This structure can provide an overall idea of the document by visualizing the relationship between the most frequent words with their most similar words. Also, constructing the \textit{Word2vec graph} requires only extracting words from the documents. Therefore, this technique can be applied to different types of documents as it only requires text. Another significant advantage is that it can be implemented across multi-lingual documents as the graph structure does not depend on the specific language.
}
This graph structure can represent both the context of the document (embedding of words associated with nodes) and the writing style of the author (word usages and co-occurrences among words).

We extract a number of features that include node/edge weight, index, degree, neighbors etc., to
capture the \textit{Word2vec graph} structure of a document. These representative sets of \textit{Word2vec graph} features with associated words allow us to perform various literary analysis tasks.
Figure~\ref{fig-example} depicts a nice visualization of the \textit{Word2vec graph} models of three different types of novels of a famous Bengali author. This example demonstrates the clear structural distinction between his three types of writing.
We are the first to identify this structural phenomenon, which can be useful in any language's literary analysis.




We evaluate the efficiency of the \textit{Word2vec graph} feature set in two major literary analysis tasks: author attribution and genre detection. 
We primarily concentrate on the literary analysis in Bengali literature since we have specific domain knowledge in this field. 
\textcolor{black}{
The major focus of this research is to show the effectiveness of the \textit{Word2vec graph} as a feature set. As a result, we do not concentrate on classification techniques. Rather, we compare the performance of the \textit{Word2vec graph} with baseline feature sets using traditional classification and clustering methods.
}
We also perform experiments in different domains, such as the editorial writings of several Bengali newspapers and a subset of English fictional writings from the Project Gutenberg corpus. 
In short, the contributions of this paper are summarized as follows. 
\begin{itemize}
    \item We present a new feature set, \textit{Word2vec graph} model, to represent any document.
    \item We utilize this graph structure and word features to solve author attribution and genre detection problems in literary analysis for various languages and domains. 
    \item We introduce two new Bengali corpus as a part of our study. We perform extensive study on these datasets along with an English literature fiction corpus. Also, we find the most contributing features from the \textit{Word2vec graph} feature set.
    
\end{itemize}

%% file: sec2_related_works.tex
\section{Related work}
\label{sec-related}

We provide a brief analysis of some stylometry analysis related tasks: author attribution and genre detection. We also address the usage of different graph-based structures in solving these tasks.

\subsection{Stylometry tasks}

\paragraph{Author attribution}
Author attribution has been extensively studied over the last few decades, with a wide range of features and classifiers.  Kjell et al.~\shortcite{kjell1994discrimination} first utilized the character bigrams and trigrams for authorship analysis. Later, the works in ~\cite{kevselj2003n,houvardas2006n} used character n-grams of different sizes and reported significant accuracy. Bag-of-words approaches have also been successful for
authorship attribution~\cite{koppel2011authorship}.
Various stylometry related features (lexical, syntactic,  semantic, application-specific) are widely used in authorship attribution for different datasets~\cite{guthrie2008unsupervised,stamatatos2009survey,neal2018surveying}. 
For classification, different machine learning classifiers have been popular, such as Decision tree~\cite{abbasi2008writeprints}, Support Vector Machine~\cite{diederich2003authorship}, Convolution Neural Network~\cite{kim2014convolutional}.

Similarly, most of the works on author attribution in Bengali language utilized the  unigram/bigram/character n-gram~\cite{das2011author, phani2015authorship}, stylometry features~\cite{islam2018authorship}, word embedding models~\cite{chowdhury2018comparative} to represent the text and applied machine learning algorithms: SVM, Neural Network to solve the classification task on the writings from newspapers/blog articles/stories.

\paragraph{Genre detection }
Several studies focus on categorizing texts based on their genre for various languages. Stamatatos et al.~\shortcite{stamatatos2000automatic} utilized the same stylometric features in the genre detection like author attribution and verification tasks for modern Greek literature. 
Word n-gram model and sentiment/emotion annotation for sentences are also utilized in predicting the genre of fictional text in several studies~\cite{amasyali2006automatic,reagan2016emotional,kar2019multi}.
Worsham and Kalita~\shortcite{worsham2018genre} presented a study on how current deep learning models compared to traditional methods perform in modern literature and discovered that merging chapters can significantly improve results for this task.


\subsection{Graph structure in literary analysis}
Representing textual documents using graphs is a well-addressed problem in NLP. Mihalcea and Tarau~\shortcite{mihalcea2004textrank} introduced TextRank, a graph-based ranking model for text processing where vertex represents text unit of various sizes (words, collections, entire sentences) and the edge represents the relation between vertices (semantic relation or contextual overlaps). TextRank is further enhanced by incorporating Word2vec~\cite{zuo2017textrank-w2v} and is utilized in keyword matching and topic extraction problems. The relationship between words at the corpus level is also explored in the word similarity graph~\cite{stanchev2014creating,feria2018constructing} and employed in the named entity recognition problem. However, these approaches formulate a representative graph from the entire corpus (collection of documents in our problem). Therefore, we can not utilize them to solve various stylometry problems  that need us to assign a unique label to each document.

Existing works from the literary domain that employ graph structures to represent stories mostly use character network graphs, where vertices symbolize characters and corresponding edges represent interactions between them~\cite{labatut2019extraction_review}. Several studies also take advantage of this to solve different stylometry tasks. Ardanuy and Sporleder~\shortcite{ardanay2015clustering} focus on social networks of characters to represent the narrative structure of novels from different genres and authors.
For each novel, they have computed a vector of literary-motivated features extracted from their network representation and performed EM clustering in terms of genres and authorship.
Our approach is different from character network analysis, since we aim to develop representative graphs from each story using text only. 

\begin{figure*}[ht]
    \centering
    \includegraphics[scale=0.4]{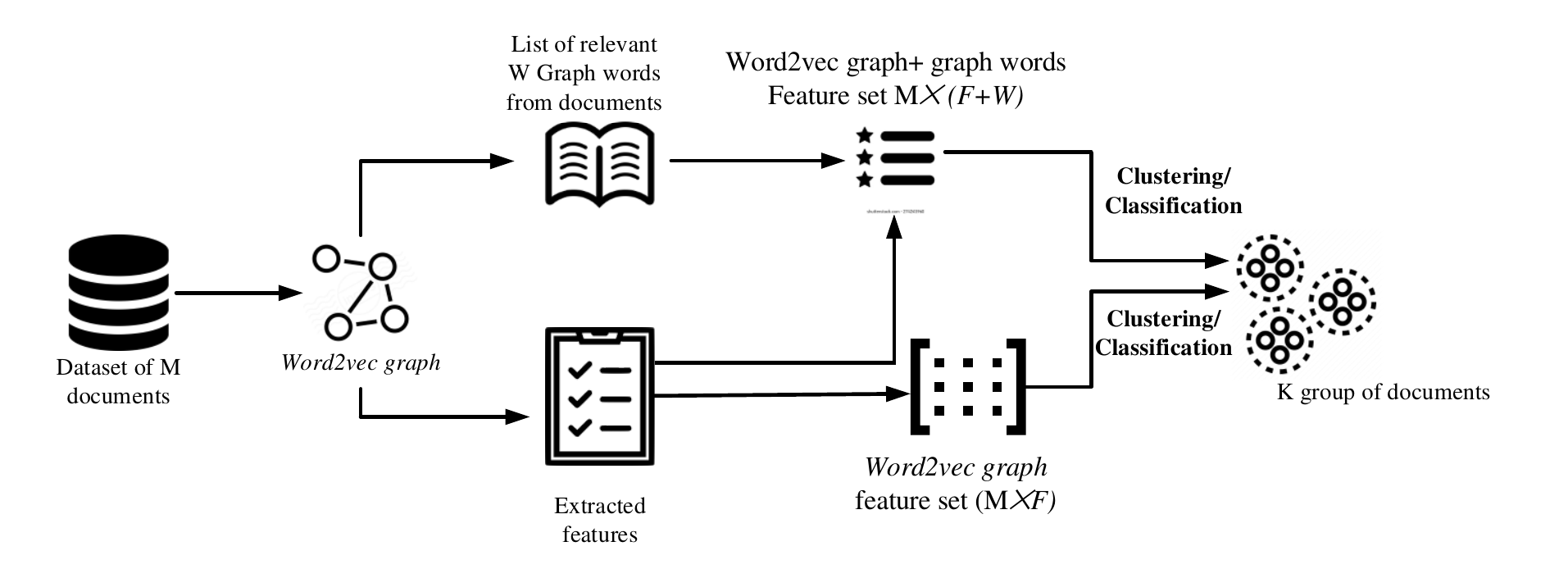}
    \caption{Overall architecture of \textit{Word2vec graph} and \textit{Word2vec graph}+graph words feature sets}
    \label{fig:architecture}
\end{figure*}

%% file: sec4_Methodology.tex
\section{The \textit{Word2vec Graph} model}
\label{sec-method_w2v}

In this section, we describe the process of constructing a \textit{Word2vec graph} from a document and then discuss the feature extraction from the \textit{Word2vec graph}. 
An overview of our proposed system is presented in Figure \ref{fig:architecture}.




\subsection{\textit{Word2vec Graph} Creation}

\textit{Word2vec graph} is developed based on the foundation of the word2vec concept in NLP.
Word2vec is a two-layer neural net, the most common word embedding technique
that represents a fixed vector size of every word in the corpus ~\cite{mikolov2013efficient}.  We utilize this word2vec embedding to represent the similarities between words in a document as a graph.
Each node in the \textit{Word2vec graph} denotes a word and edge between nodes represent their relation in the document. We consider edge weight as the cosine similarity between vector representations of corresponding words. 
The concept of \textit{Word2vec graph} is motivated from the word similarity graph~\cite{stanchev2014creating,feria2018constructing}.
   In their work, a single word similarity graph is created for the whole corpus and the similarity between words is measured by their co-occurrence.
 However, in our problem, we represent each document as a \textit{Word2vec graph}, and
the similarity between words is measured from the word2vec vector representations in the embedding space.
Word similarity graph is commonly used for community detection problems to find related groups of words/synsets, where \textit{Word2vec graph} is used to classify a document in our study. 
\color{black}

To generate a \textit{Word2vec graph} from a sample document, we first train a word2vec model from the tokenized format of the document. This model transform each word $w$ in the document to a dense vector $\bar w$ with $l$ dimension. We take top $N$ words (\textit{core} words) in the document with most frequency. For each \textit{core} word $i$, we consider $K$ most similar words in the documents.  We calculate the cosine similarity between each of these similar words $j$ and corresponding \textit{core} word $i$, which represents the edge weight $\omega_{ij}$. We also consider each node weight $w_i$, as the relative frequency of that word in the document.
\textcolor{black}{We categorize nodes in the \textit{Word2vec graph} into three types according to their relative index based on frequency in document and connectivity with \textit{core} nodes. Similarly, edges are also classified by their connectivity to different types of nodes.} 

\begin{itemize}
    \item \textit{\textbf{Core nodes, edges:}} The nodes representing the most frequent top-$N$ words in the document. Any edge among \textit{core} nodes is represented as the \textit{core edge}.
    
    \item \textit{\textbf{Multi nodes, edges:}} 
    The nodes which are connected to two or more \textit{core} nodes in the graph. The edge between a \textit{multi} node and \textit{core} node is represented as \textit{multi edge}.
    
    \item \textit{\textbf{Boundary nodes, edges:}} 
    The nodes which are connected to only one \textit{core} node. The edge between a \textit{boundary} node and \textit{core} node is represented as \textit{boundary edge}. 
\end{itemize}

\textcolor{black}{
The primary intention of this structure is to represent the document using the document's most frequent words (\textit{core} nodes). The words (\textit{multi}/\textit{boundary} nodes) related to the most frequent words  represent the context of the documents (for example, in newspaper articles, they can represent keywords about what the article is). These are usually personal nouns and verbs that identify the primary characters and the most prevalent actions described in the document. The way in which these words are related to the most frequent words can disclose context and author style. Words related to several most frequent words (\textit{multi} nodes) show a central theme, but words related to only one most frequent word (\textit{boundary} nodes) exhibit the author's personal style. Therefore, it motivates us to divide the nodes  into three categories based on connectivity. 
}
 
While creating \textit{Word2vec graph}, we also address the effect of function words/stopwords since they play a significant role  in various NLP tasks~\cite{argamon2005measuring}. 
Therefore, we consider both versions of \textit{Word2vec graph}, one \textit{with-stopwords}, and other \textit{without-stopwords} for each document.
We do not perform any other pre-processing tasks, such as lemmatization/stemming, to keep the intact form of words.
The choice of the value of $N, K$, and other parameters are discussed in Subsection \ref{subsec-exp_setup}. 


\begin{table*}[h]
\centering
\begin{tabular}{|l|c|}
\hline
\textbf{Feature   }                                                                            & \textbf{\#}    \\
\hline
No. of \textit{core}/\textit{multi}/\textit{boundary} nodes/edges                                             & 6     \\
Min/max/avg/sum of \textit{core}/\textit{multi}/\textit{boundary} nodes/edges weights                             & 18    \\
Count/min/max/avg of \textit{core} nodes degree (considering \textit{core}/\textit{multi}/\textit{boundary} edge only) & 10    \\
Degree count of \textit{core} nodes (considering \textit{core}/multiple edge only)                      & 30/40 \\
Min/max/avg/stdv index of \textit{multi}/\textit{boundary} nodes                                     & 8     \\
No. of \textit{multi}/\textit{boundary} nodes index under threshold                                  & 2     \\
\textit{core} nodes having degree 0-5 (considering \textit{multi} edges only)                        & 6     \\
No. of \textit{core} nodes having degree equal to min/max/greater than avg/smaller than avg           & 4  \\ 
\hline
\textbf{Total features}          & \textbf{85/95}  \\ 
\hline
\end{tabular}
\caption{Extracted features from \textit{Word2vec graph}}
\label{tab-w2v_feature}
\end{table*}

\subsubsection{Feature Extraction from graph}
\textcolor{black}{Different graph embedding methods have been popular recently to represent the graph structure, such as DeepWalk~\cite{perozzi2014deepwalk}, Graph Convolution Network~\cite{kipf2016semi}, Graph2vec~\cite{narayanan2017graph2vec}. However, the distinction between \textit{Word2vec graph} structures can be interpreted from the count, connectivity of different types of nodes as depicted in Figure 1. For example, the number of \textit{core} nodes in a \textit{Word2vec graph} is constant (N), whereas the number of \textit{multi} \& \textit{boundary} nodes varies across documents. Therefore, we extract some widely utilized graph statistics regarding three types of nodes \& edges as \textit{Word2vec graph} features that are intuitively meaningful for classification. }

Apart from the count, distribution of weight, degree of various nodes/edges, we also consider the relative index of nodes (ordered index of words based on frequency). 
\textcolor{black}{Because we intend to use the frequency of associated words with \textit{multi}/\textit{boundary} nodes that are linked to the most frequent words.}
There is a total of 95 and 85 features respectively, representing the graph structure for \textit{with-stopwords} and \textit{without-stopwords} version. 
Table~\ref{tab-w2v_feature} provides a list of features that we extract from \textit{Word2vec graph} in both  versions.
Given a dataset of $M$ documents, we extract relevant features from the graphs to convert the dataset as a $M \times F$ matrix, where $F$ is the total number of features in the \textit{Word2vec graph} feature set. 

\textcolor{black}{
However, this feature list is devoid of any actual word-related information. This is due to the fact that word indexes differ for each document, and we are not utilizing the information about which nodes represent which words. As a result, it might miss some contextual information. Therefore, we utilize combinations of the \textit{core}, \textit{multi}, and \textit{boundary} nodes associated words set (size $W$) as graph words. The graph words, along with the \textit{Word2vec graph} features, constitute the  \textit{Word2vec graph+graph words (w2v+)} feature set. 
}

%% file: sec5_experiment.tex
\section{Experimental study}
In this section, we evaluate the performance of \textit{Word2vec graph} feature set and compare it with the baseline approaches. We present an overview of corpus creation, then go over the baseline feature sets before presenting the comprehensive evaluation results.
We also identify the most prominent attributes in \textit{Word2vec graph} feature set that contribute significantly to various tasks.
Our code and data are publicly available\footnote{\url{https://cutt.ly/svLjSgk}}.

\input{sec3_corpus_creation}

\input{sec_baseline}



\subsection{Experimental setup}
\label{subsec-exp_setup}
\paragraph{Parameter values: }
\begin{figure}
    \centering
    \includegraphics[scale=0.65]{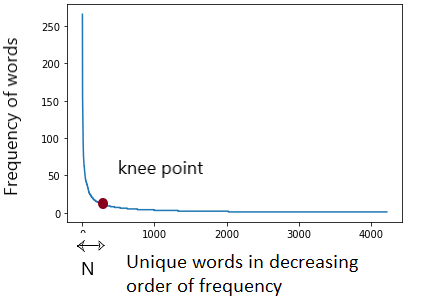}
    \caption{Avg. distribution of frequency of unique words in the documents (Bengali literature corpus)}
    \label{fig:distribution-words}
\end{figure}


We experiment with distinct values of $N, K$ and different variations of \textit{Word2vec graph}. For example, we consider the most similar words of the \textit{multi} nodes (like \textit{core} nodes), thus, introducing a new layer of nodes and extending the network. 
However, the performance does not increase, and we keep the same architecture (described in Section 3). 
The frequency of unique words in a document follows an exponentially decreasing distribution in all cases (Figure \ref{fig:distribution-words}). 
We calculate the average knee point in the distribution curve, and the corresponding number of words are considered \textit{core} words since words beyond this knee point do not have a substantial frequency. 
We set  $N=20$ and $N=15$ for \textit{with-stopwords} and \textit{without-stopwords} version respectively.
\textcolor{black}{
Increasing the value of $N$ makes the core of the graph denser (the number of \textit{core} nodes and \textit{core} edges increases), and thus we miss multi words information. Also, decreasing the value of $N$ will result in the loss of \textit{core} words information (context) from the document.
}

\textcolor{black}{
We consider w2v dimension $l=100$ since it should be sufficient to calculate the similarity between words and make the text processing faster. Increasing $l$ has little effect on the word similarity matrix, and our \textit{Word2vec graphs} will remain unchanged. We experiment with different values of $K$ of magnitude $\mathcal{O}(N)$ and $\mathcal{O}(N^2)$  to find the optimal value. It seems that the visual distinction between the \textit{Word2vec graphs} (as well as the average feature set difference between various classes) is maximized when the total nodes count is $\mathcal{O}(N^2)$. Therefore, it motivates us to experiment with the value of $K$ in the same range as $N$. We use $K= 10$ in all cases since it is sufficient to preserve the central theme and style of authors (from \textit{multi} \& \textit{boundary} nodes). Increasing $K$ makes the graph more sparse, and thus removes the second layer (\textit{multi} nodes). Decreasing $K$ eliminates the outer layer (\textit{boundary} nodes) from the \textit{Word2vec graph}. Slight variations in the values of $N, K$ do not have the noticeable impact on the result.
}


\paragraph{Performance metric:}
For clustering, we utilize the Hungarian algorithm~\cite{jonker1986hungarian} to obtain an optimum allocation of the stories from the predicted cluster label and the original label.
Since there is a class imbalance in the datasets, we adopt weighted F1 score as the performance metric for SVM and k-means. 

\begin{table}[]
\begin{tabular}{|l|c|c|c|c|}
\hline
\multicolumn{1}{|c|}{\multirow{2}{*}{\textbf{Feature set}}}       & \multicolumn{2}{c|}{\textbf{Author}} & \multicolumn{2}{c|}{\textbf{Genre}} \\ \cline{2-5} 
\multicolumn{1}{|c|}{}                                            & \textbf{A1}             & \textbf{A2}          & \textbf{A1}           & \textbf{A2}         \\ \hline
w2v graph                                                         & 75.28                    & 51.11                 & 87.14                  & 68.89                \\ \hline
\begin{tabular}[c]{@{}l@{}}w2v graph+\\  graph words\end{tabular} & 81.40                    & 63.88                 & \textbf{92.30}         & 69.62                \\ \hline
TF-IDF                                                            & 43.62                    & 57.94                 & 85.43                  & 50.43                \\ \hline
\begin{tabular}[c]{@{}l@{}}TF-IDF (top \\ 100 words)\end{tabular} & 66.34                    & 56.91                 & 88.83                  & 55.97                \\ \hline
Stylometry                                                        & 52.65                    & 29.22                 & 87.40                  & 85.99                \\ \hline
\begin{tabular}[c]{@{}l@{}}Character \\ ngrams \\ (C-ngram)\end{tabular} & 56.34                    & 46.91                 & 78.83                  & 55.97                \\ \hline
\begin{tabular}[c]{@{}l@{}}Multilingual\\  BERT\end{tabular}      & \textbf{85.98}           & 41.47                 & 62.22                  & 44.74                \\ \hline
\begin{tabular}[c]{@{}l@{}}AVG of w2v \\ vectors\end{tabular}     & 72.49                    & 41.47                 & 78.40                  & 47.65                \\ \hline
\begin{tabular}[c]{@{}l@{}}w2v graph \\ words only\end{tabular}   &  48.13                        &       56.14                & 46.33                  & 52.01                \\ \hline
\end{tabular}
\caption{Performance of different feature sets. A1: SVM classification, A2: k-means clustering }
\label{Table-overall_result}
\end{table}

\begin{table*}[ht]
\centering
\begin{tabular}{|l|c|c|c|c|c|c|l|}
\hline
\textbf{Criteria}                    & \textbf{\#} & \textbf{w2v} & \textbf{w2v+} & \multicolumn{1}{l|}{\textbf{m-BERT}} & \textbf{TF-IDF} & \textbf{Stylo} & \textbf{C-ngram} \\ \hline
Science fiction (HM, ZI)             & 43          & \textbf{1}            & \textbf{1}    & \textbf{1}                           & \textbf{1}               & .887           & .915             \\ \hline
Teenage novel (SM, ZI)               & 50          & .916         & .978          & \textbf{1}                           & .911            & .879           & .911             \\ \hline
Long historical novel (HM, SG, SM)   & 16          & .838         & \textbf{1}    & \textbf{1}                           & .815            & .789           & .875             \\ \hline
Series novel (HM, SG)                & 79          & .738         & \textbf{1}    & \textbf{1}                           & \textbf{1}      & .713           & \textbf{1}       \\ \hline
Thriller (SG, SM)                    & 43          & .84          & \textbf{1}    & \textbf{1}                           & .979            & .761           & \textbf{1}       \\ \hline
Short story (HM, RT, SC, SMM)        & 308         & .457         & .574          & .589                                 & \textbf{.785}   & .343           & .715             \\ \hline
Novel (RT, BC, SC)                   & 50          & .512         & .915          & \textbf{1}                           & .915            & .416           & .825             \\ \hline
Novel (SG, SM, SMM)                  & 20          & .691         & .801          & \textbf{1}                           & .875            & .463           & .767             \\ \hline
Novel, short story (HM, RT, SC, SMM) & 462         & .482         & .567          & .606                                 & .615            & .397           & \textbf{.635}    \\ \hline
Novel, historical novel (HM, SG, SM) & 131         & .780         & .844          & \textbf{.932}                        & .921            & .567           & .874             \\ \hline
\end{tabular}
\caption{Weighted F1 score for various feature sets (\textit{without-stopwords} version) on author attribution task in Bengali literature corpus using SVM classification.}
\label{Tab-author_result}
\end{table*}

\begin{figure*}[h]
	\centering
	\begin{subfigure}{\columnwidth}
		\centering
		\includegraphics[width=0.7\columnwidth]{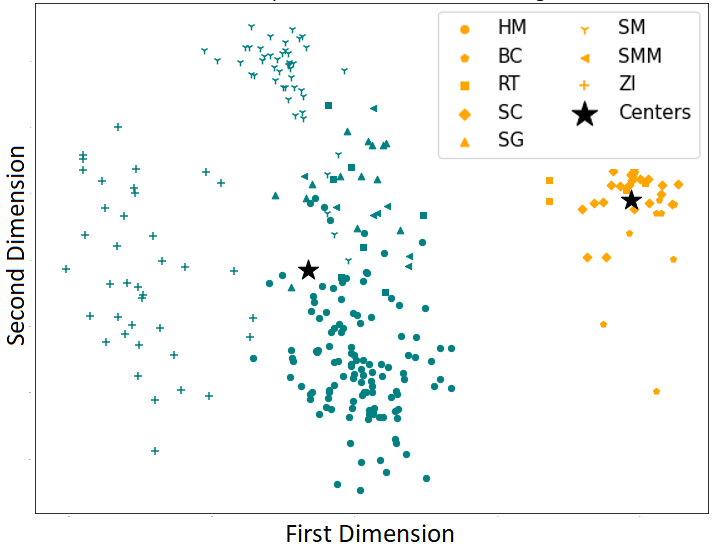}
		\caption{Two clusters}
	\end{subfigure}%
	\begin{subfigure}{\columnwidth}
		\centering
		\includegraphics[width=0.7\columnwidth]{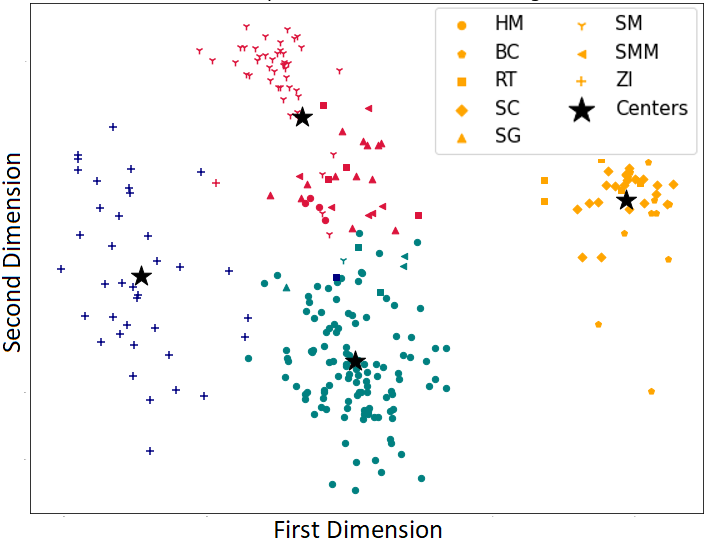}
		\caption{Four clusters}
	\end{subfigure}%
	\caption{k-means clustering visualization for all authors using \textit{Word2vec graph} feature set  \textit{without-stopwords} version. Every point denotes a sample story. Each color represents a cluster and each marker represents the writing of an author.}
	\label{fig-tf_idf(with)}
\end{figure*}

\subsection{Result analysis}

\textcolor{black}{
In Table \ref{Table-overall_result}, we examine the performance of all feature sets in author attribution (8 authors) and genre detection (5 genres) tasks using the Bengali literature dataset.
For genre detection, we consider writings from five genres: novel, short story, long/historical novel, thriller \& science fiction since other genres have a limited number of writings in total and do not contain writings from more than one author. 
The \textit{without-stopwords} version performs significantly better in all feature sets, and so we report it in Table \ref{Table-overall_result}.
}

\textcolor{black}{
Multilingual BERT (m-BERT) provides the highest F1 score in author attribution. However, when m-BERT is trained on genre detection task, it does not perform well. \textit{Word2vec+graph word} feature set provides the highest performance in genre detection, and its performance in author attribution is also satisfactory. The \textit{Word2vec+graph word} outperforms TF-IDF, TF-IDF with top 100 words, \textit{Word2vec graph} words only, Avg. of Word2vec vectors. Therefore, \textit{Word2vec graph} can portray author/genre related information through its structure, where other feature sets trained on specific instances performed well in either author or genre classification. 
}

\textcolor{black}{
In most cases, classification achieves higher performance than clustering since the training dataset incorporates specific information regarding author/genre. However, clustering often derives a natural separation of the writings irrespective of their genre/author, combining other factors, such as dialect, author origin, or time. We discuss these phenomena in detail in subsequent subsections. Also, we observe that feature set, such as TF-IDF with top 100 words or \textit{Word2vec graph} words does not perform well. Therefore, it indicates the suitability of \textit{Word2vec graph} based structural features in representing literary text even if having a significantly lower number of features or less resource requirement than TF-IDF or m-BERT. We provide corpus specific detailed performance analysis of different feature sets in the following subsections. 
}


\subsubsection{Bengali literature corpus}

\begin{table*}[ht]
\centering
\begin{tabular}{|l|c|c|c|l|c|c|l|}
\hline
\textbf{Criteria}                     & \textbf{\#} & \textbf{w2v}  & \textbf{w2v+} & \textbf{m-BERT} & \textbf{TF-IDF} & \textbf{Stylo} & \textbf{C-ngram} \\ \hline
HM, RT, SC, SMM (novel, short story)  & 462         & .794          & .794          & .575            & .626            & \textbf{.805}  & .781             \\ \hline
HM, SG, SMM (novel, historical novel) & 131         & \textbf{.936} & \textbf{.936} & .616            & .676            & .785           & .699             \\ \hline
HM (novel, Himu, Misir Ali)           & 146         & .431          & .664          & .615            & \textbf{.893}   & .519           & .871             \\ \hline
HM (novel, science fiction)           & 113         & .751          & \textbf{.782} & .751            & .689            & .698           & .698             \\ \hline
SM (teenage novel, thriller)          & 46          & .767          & \textbf{.962} & .825            & .825            & .767           & .791             \\ \hline
ZI (teenage novel, science fiction)   & 44          & \textbf{.977} & \textbf{.977} & .615            & .583            & .646           & .636             \\ \hline
\end{tabular}
\caption{Weighted F1 score for various feature sets (without-stopwords version) on genre detection task in Bengali literature corpus using SVM classification.}
\label{Tab-genre_result}
\end{table*}

\begin{table*}[h]
\centering
\begin{tabular}{|l|c|l|c|c|l|c|c|l|}
\hline
\textbf{Critera}              & \textbf{\#} & \textbf{Class \#} & \textbf{w2v} & \textbf{w2v+} & \textbf{m-BERT} & \textbf{TF-IDF} & \textbf{Stylo} & C-ngram       \\ \hline
Detective and mystery stories & 52          & 2                 & .856         & \textbf{1}    & \textbf{1}      & .924            & .878           & .875          \\ \hline
Love stories                  & 17          & 2                 & .646         & \textbf{1}    & \textbf{1}      & \textbf{1}      & .684           & .715          \\ \hline
Historical fiction            & 19          & 2                 & .578         & \textbf{1}    & \textbf{1}      & .947            & .526           & .875          \\ \hline
Adventure stories             & 39          & 4                 & .435         & .861          & \textbf{.937}   & .713            & .461           & .713          \\ \hline
Western stories               & 27          & 2                 & .782         & .712          & \textbf{.812}   & \textbf{.812}   & .769           & \textbf{.812} \\ \hline
Science fiction               & 28          & 3                 & .720         & \textbf{.854} & \textbf{.854}   & .776            & .758           & .776          \\ \hline
All genres                    & 201         & 6                 & .504         & .681          & \textbf{.735}   & .696            & .453           & .716          \\ \hline
\end{tabular}
\caption{Classification result on author attribution in English literature fiction corpus}
\label{tab-author_guten}
\end{table*}

\paragraph{Author attribution (Table \ref{Tab-author_result}): }

\textcolor{black}{
Multilingual BERT feature set achieves the highest F1 score in nearly all genres. \textit{w2v+} performs almost equal to m-BERT. 
Although TF-IDF feature set shows poor performance in overall author attribution, it performs well in specific criteria where the sample size is small. 
}
\textcolor{black}{
Both m-BERT and \textit{w2v+} perform poorly when short stories are involved because of the limited amount of text. TF-IDF and Character n-gram feature sets perform better in these cases. It is noticeable that the performance of \textit{w2v+} is better for specific genres, such as science fiction, teenage novel, and thriller. \textit{core} words and their associated words can capture the author's style and choice of words effectively in the stories of these genres. 
We observe several patterns in the \textit{Word2vec graph} structure during manual evaluation. For example, there are fewer \textit{boundary} nodes in the novels of Humayun Ahmed (HM) than in Rabindranath (RT), which means that there are more unique words in his (RT) writings that are related to the most frequent words.
}

\textcolor{black}{
We also experiment with k-means clustering within these criteria, and the performance is lower than classification. However, clustering using \textit{Word2vec graph} feature set often reveals significant patterns and a natural separation among writings. Figure~\ref{fig-tf_idf(with)} shows the clustering visualization on the total corpus (except short stories since they all form a separate cluster) with a fixed number of clusters (\textit{without-stopwords} version). If we consider two clusters, writings of Bankim, Sarat, Rabindranath (partially) form one cluster (dialect: \textit{Sadhu}). The second cluster constitutes of rest of the writings of Rabindranath and other authors (dialect: \textit{Cholito}). If we increase the number of clusters (for example, four), writings of Humayun Ahmed and Muhammed Zafar Iqbal form their separate clusters (Bangladeshi authors), and writings of Sunil, Shirshendu, Samaresh, and partially Rabindranath (West Bengal and previous Bengali authors who wrote in \textit{Cholito}) constitute a single cluster. We cannot infer these patterns from the direct classification of other feature sets.
}

\paragraph{Genre detection (Table~\ref{Tab-genre_result}):}
\textit{Word2vec graph} and \textit{w2v+} performs better in almost all cases. 
As we have presented earlier, m-BERT representation of writings does not perform well in the genre detection task.
TF-IDF and character n-gram feature sets are successful, particularly in distinguishing series writings (\textit{Himu} and \textit{Misir Ali} series) of Humayun Ahmed from normal novels. Each series contains the same set of characters in different books, and so word/character usage follows specific patterns. Stylometry feature set performs the best for short stories since it has some specific stylometry features, such as total sentences/ words that make it easily separable from other genres. All feature sets perform better in \textit{without-stopwords} version since stopwords mostly do not contain any genre-related information.

\begin{table*}[h]
\centering
\begin{tabular}{|l|c|c|c|l|c|c|l|}
\hline
\textbf{Criteria}                                                                       & \textbf{\#} & \textbf{w2v}  & \textbf{w2v+} & \textbf{m-BERT} & \textbf{TF-IDF} & \textbf{Stylo} & \textbf{C-ngram} \\ \hline
All authors                                                                             & 334         & \textbf{.565} & .537          & .497            & .457            & .541           & .447             \\ \hline
\begin{tabular}[c]{@{}l@{}}All authors with\\ book count \textgreater{}=10\end{tabular} & 192         & .591          & \textbf{.632} & .625            & .616            & .596           & .596             \\ \hline
\end{tabular}
\caption{Classification result on genre detection in English literature fiction corpus}
\label{tab-genrer_guten}
\end{table*}

\begin{figure*}[ht]
	\centering
	\begin{subfigure}{0.66\columnwidth}
		\centering
		\includegraphics[height=3.5cm]{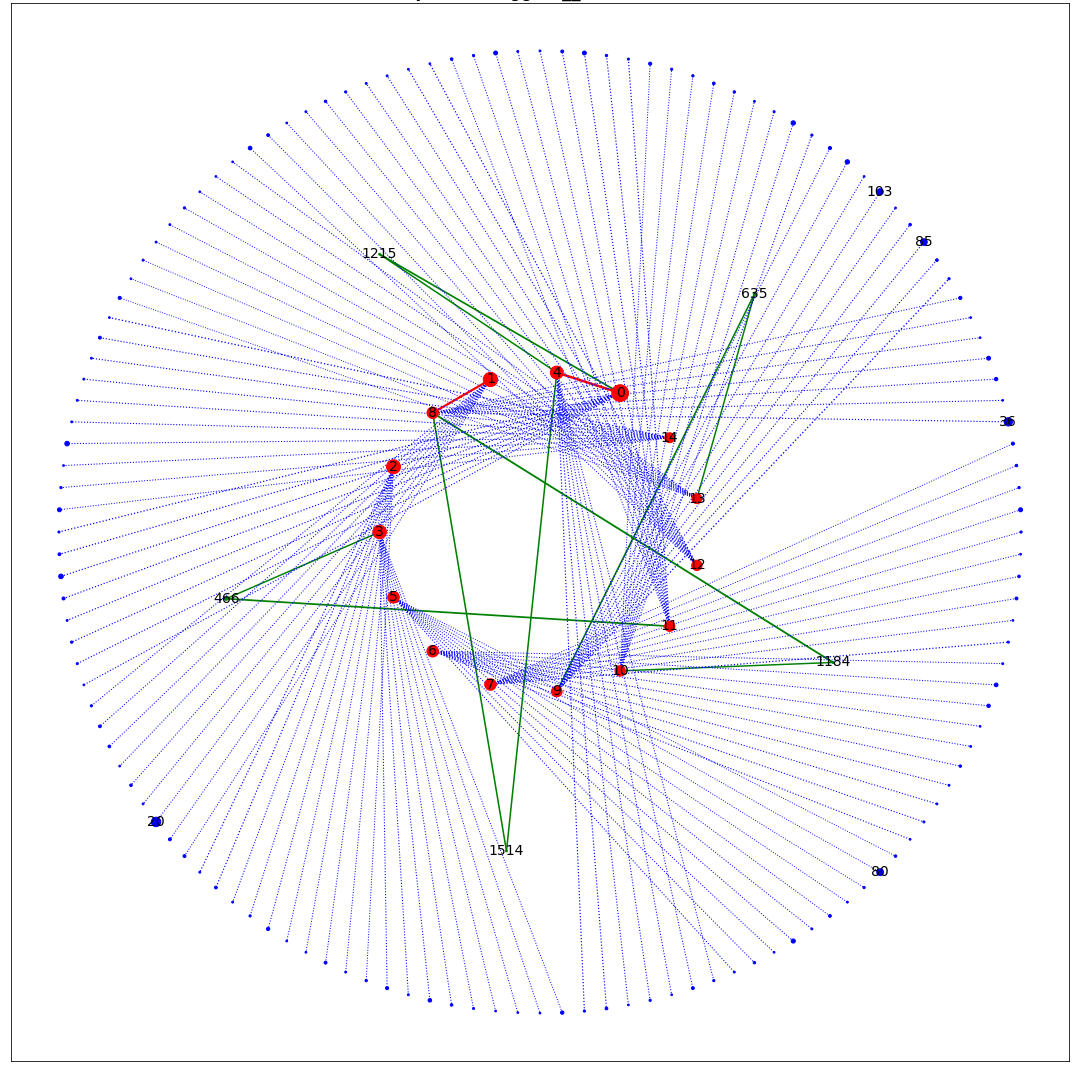}
		\caption{Historical fiction}
		\label{subfig-his}
	\end{subfigure}%
	\begin{subfigure}{0.66\columnwidth}
		\centering
		\includegraphics[height=3.5cm]{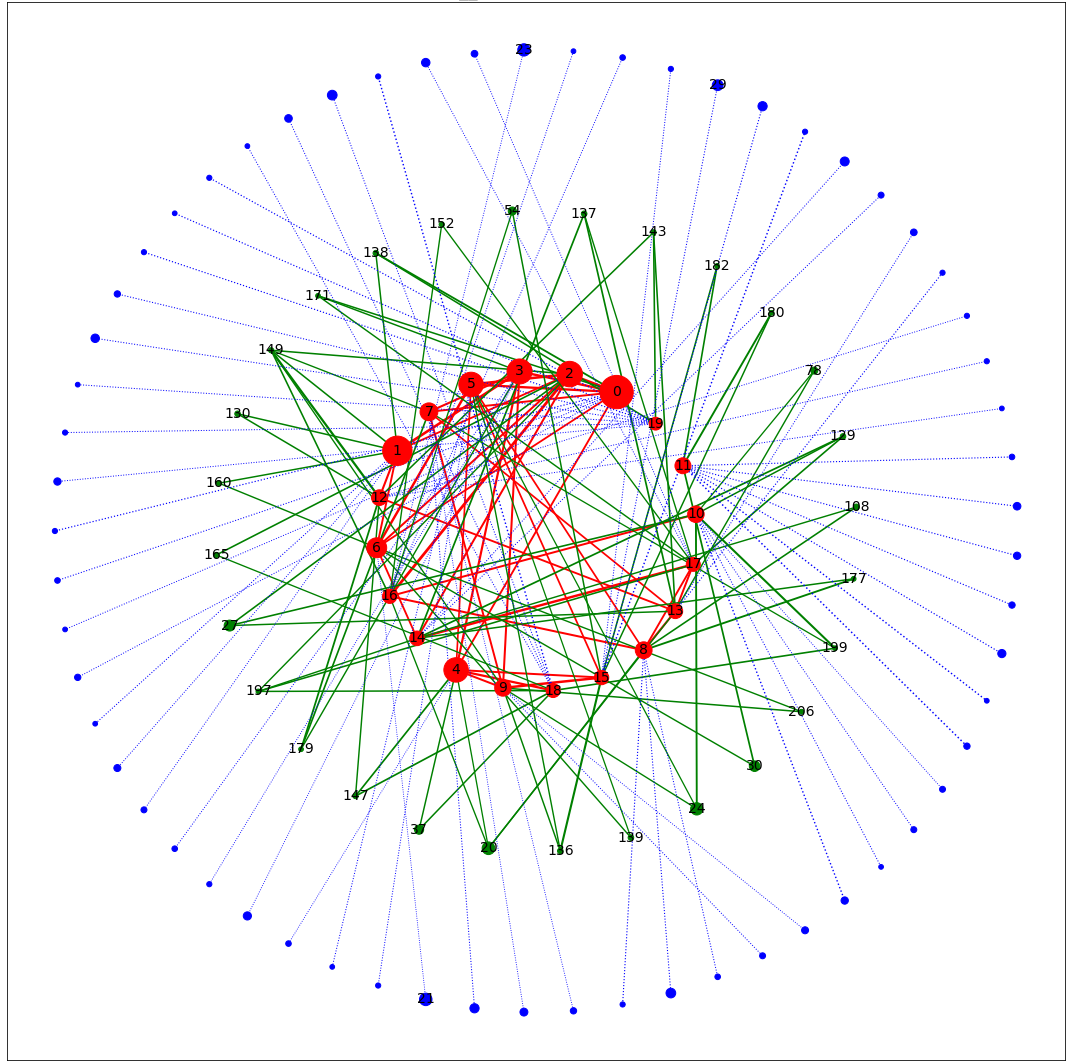}
		\caption{Detective and mystery stories}
		\label{subfig-love}
	\end{subfigure}%
	\begin{subfigure}{0.66\columnwidth}
		\centering
		\includegraphics[height=3.5cm]{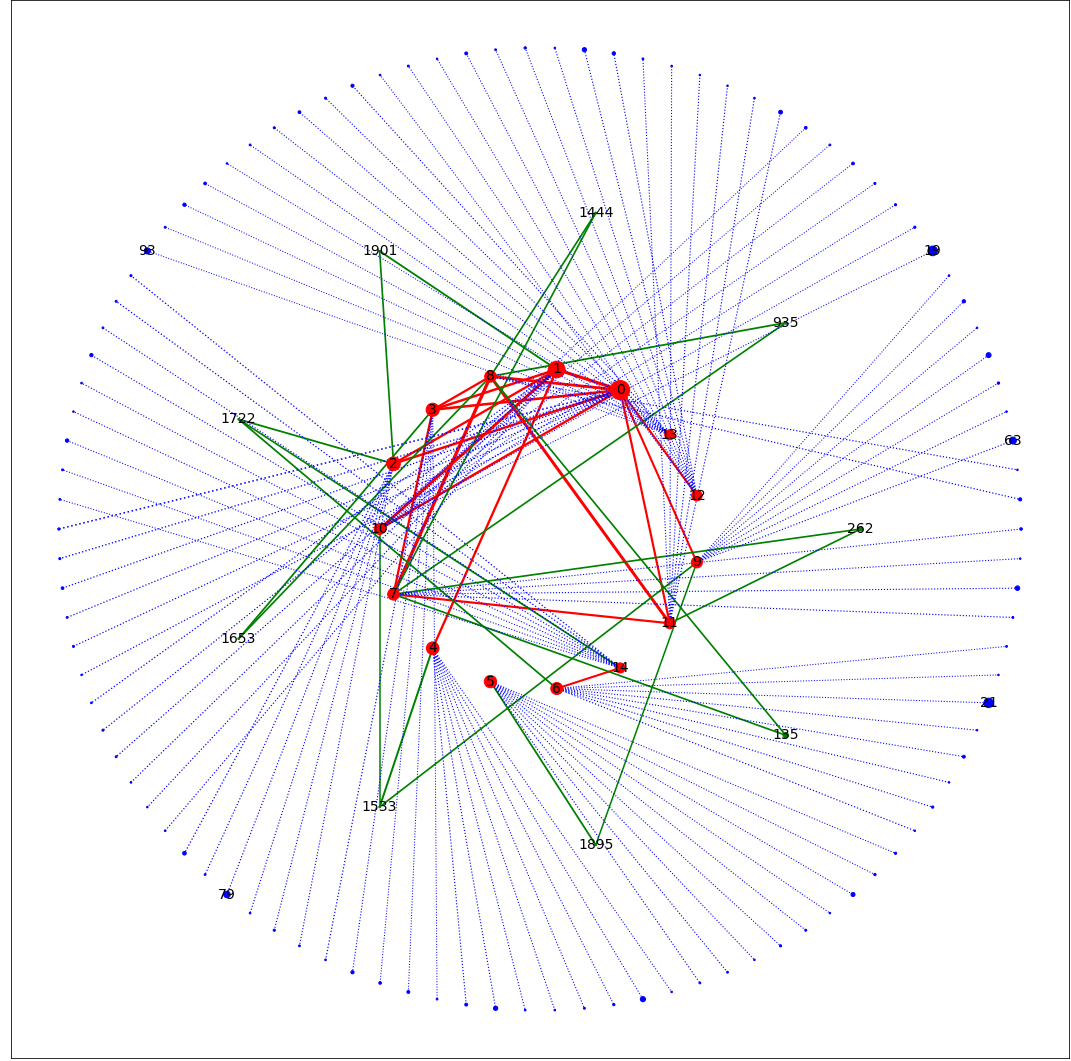}
		\caption{Love stories}
		\label{subfig-detec}
	\end{subfigure}%
	\caption{Average \textit{Word2vec graph} representation of different genre novels in English literature dataset
	}
	\label{fig-example_eng}
\end{figure*}


\subsubsection{English literature fiction corpus}

\paragraph{Author Attribution (Table \ref{tab-author_guten}):}
As discussed earlier, m-BERT achieves the highest performance in author attribution. Our method \textit{w2v+}  delivers nearly similar performance in most scenarios. 
The \textit{without-stopwords} version works better in all cases, and so we report its performance in the table \ref{tab-author_guten}. 
TF-IDF and character n-gram feature sets perform better, particularly in western story genres. Authors of these books often use specific, localized, regional words. Therefore, the limited number of words considered in \textit{w2v+} feature set can not capture them all.

\paragraph{Genre Detection (Table \ref{tab-genrer_guten}):}
Interestingly, the \textit{Word2vec graph} structure performs best if we consider all authors. \textit{w2v+} performs best after removing the outliers. 
Both are better than the baseline feature sets. Unlike author attribution, m-BERT does not provide the highest performance in genre detection. Because this corpus contains writings from a varied period, word choices align with the author's style. Therefore, TF-IDf and character n-gram also perform poorly in genre detection.

\textcolor{black}{
We observe some interesting patterns in the \textit{Word2vec graph} structure of different genres as portrayed in Figure~\ref{fig-example_eng}. Overall, historical fiction includes fewer \textit{multi} nodes \& edges and higher \textit{boundary} nodes. It indicates more unique words used by authors in these genres, and most frequent words (\textit{core} nodes) often do not share similarities with these words in embedding space.  \textit{Word2vec graph} from detective novels contain higher \textit{core} edges \& \textit{multi} nodes since content specific words have a high frequency in these genre. However, we observe similar \textit{core} edges and fewer \textit{multi} nodes in love genres than in detective novels.
}

\begin{table*}[h]
\centering
\begin{tabular}{|l|c|c|c|l|c|c|l|}
\hline
\textbf{Criteria}                  & \textbf{\#} & \textbf{w2v}  & \textbf{w2v+} & \textbf{m-BERT} & \textbf{TF-IDF} & \textbf{Stylo} & \textbf{C-ngram} \\ \hline
3 different newspaper (Pr, It, An) & 765         & .455          & .786          & \textbf{.831}   & .762            & .416           & .701             \\ \hline
All Bangladeshi (It, Pr, Ju,In)    & 730         & .572          & \textbf{.619} & .604            & .604            & .517           & .596             \\ \hline
Bangladeshi cholito (Pr, In, Ju)   & 678         & \textbf{.691} & .526          & .614            & .440            & .634           & .481             \\ \hline
All cholito (An, Pr, In, Ju)       & 935         & .655          & \textbf{.701} & \textbf{.701}   & .553            & .507           & .515             \\ \hline
All 5 newspaper                    & 1189        & \textbf{.578} & .491          & .567            & .482            & .531           & .505             \\ \hline
\end{tabular}
\caption{Classification performance on newspaper editorial corpus (\textit{with-stopwords} version)}
\label{Tab-Newspaper_result}
\end{table*}

\subsubsection{Newspaper editorial corpus (Table \ref{Tab-Newspaper_result}):}
In most cases, \textit{w2v+} or \textit{Word2vec graph} feature set achieve highest performance.
\textcolor{black}{Multilingual BERT performs better when the west Bengal newspaper Anandabazar is present since its context is different from others.}
Likewise, TF-IDF only performs better with the inclusion of \textit{Sadhu-Bhasha} newspaper articles (Ittefaq). 
Unlike other datasets, \textit{with-stopwords} version performs better since editorial articles are short, and removing the stopwords can reduce the available information. 
\textcolor{black}{\textit{Word2vec graph} structure of different newspapers portrays headliner words as its \textit{core} nodes and technical key/context words as \textit{multi} nodes. }

 \begin{figure*}[h]
	\centering
	\begin{subfigure}{0.66\columnwidth}
		\centering
		\includegraphics[width=0.9\columnwidth]{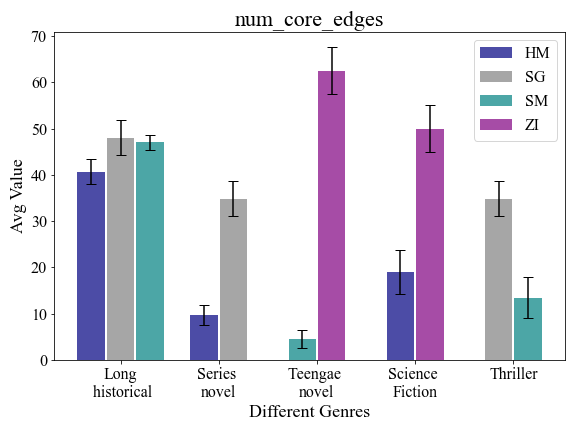}
		\caption{Number of \textit{core} edges}
	\end{subfigure}%
	\begin{subfigure}{0.66\columnwidth}
		\centering
		\includegraphics[width=0.9\columnwidth]{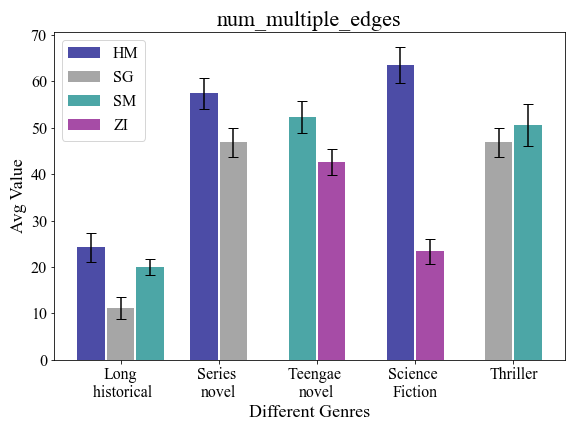}
		\caption{Number of \textit{multi} edges}
	\end{subfigure}%
	\begin{subfigure}{0.66\columnwidth}
		\centering
		\includegraphics[width=0.9\columnwidth]{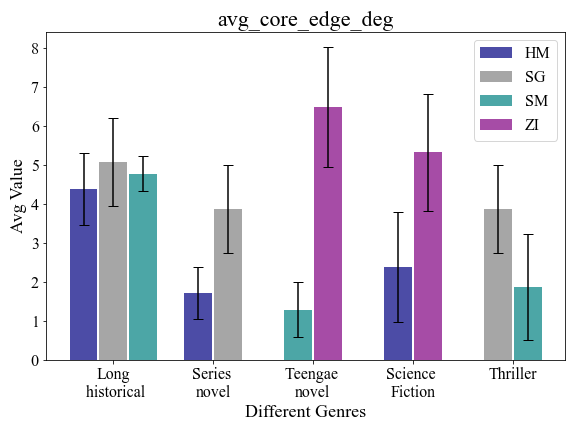}
		\caption{Average degree of \textit{core} nodes}
	\end{subfigure}%
	\caption{Avg. value (with SD marked) of some features in \textit{Word2vec graph} for different domains.}
	\label{fig-w2v_other}
\end{figure*}

\subsection{Most contributing features selection}

We quantify the usefulness of different features by their discriminative power to tell classes apart~\cite{alelyani2018feature}. We apply the method in~\cite{burns2008anova} to examine the means for each cluster/class on each dimension using analysis of variance (ANOVA) to assess how distinct they are. The magnitude of the F values obtained on each dimension indicates how well that dimension discriminates between classes. We also validate these findings by sample chi-square test and tree-based feature selection method~\cite{ghaemi2016tree}.


No. of \textit{core} edges plays an important feature in all tasks since it demonstrates the relationship between most frequent words of the story. Min/Max/Avg of \textit{multi/boundary} edge index/weight/degree also works as significant attributes for author identification in particular genres. For example, Humayun Ahmed science fictions show a higher number of \textit{multi} edges than Zafar Iqbal, and their average index is also higher. It indicates that most frequent words have higher similarity with words with lower index (more frequent) than words with higher index (less frequent). We present some examples  for author attribution in different genres in Figure~\ref{fig-w2v_other}. 
In differentiating short stories from other genres, No. of \textit{core} edges, \textit{boundary} nodes, \textit{core} nodes having \textit{core} edges are  significant features since the amount of text and unique words is limited here.

\subsection{Discussion, limitation and future study }

\textcolor{black}{
Our primary contribution is to propose a novel \textit{Word2vec graph}  based feature extraction technique for a text document.. 
It can also visualize the structural distinction between different documents from a new perspective. For example, \textit{core} nodes in  \textit{Word2vec graph} of stories indicate the most frequently used verbs/adjectives, and \textit{multi} nodes represent the central character names or context words. In newspaper editorial corpus, \textit{core} \& \textit{multi} nodes represent headline-related words and technical words, respectively. 
Our proposed feature set achieves better performance than TF-IDF \& stylometry feature sets, even if having a significantly lower number of features. It provides almost similar performance to m-BERT representation of documents as a feature set in author attribution.
We have shown that considering graph words information increases the performance of  \textit{Word2vec graph} approach in most cases.
}


There are some limitations to our study. Since the major focus of our research is to introduce a new feature set, we employ a more simplistic classification and clustering scheme in our tasks.
Therefore, we would like to extend the volume of our experiments,
utilize graph embeddings to extract features from \textit{Word2vec graph} and perform other literary tasks.  


%% file: sec3_corpus_creation.tex
\subsection{Corpus creation}
\label{sec-corpus}

\paragraph{Bengali literature corpus: }

Bengali exhibits two major forms of dialects, and thus two styles of writing have emerged that involve somewhat different vocabularies and syntax~\cite{islam2003banglapedia}. These are \textit{Sadhu-Bhasha} and \textit{Cholito-Bhasha}.
\textit{Sadhu-Bhasha} or chaste language is an old written formal style of  Bengali language (widely used in literature up to the first half of the 19th century), with longer verb inflections and more of a Pali/Sanskrit-derived vocabulary. \textit{Cholito-Bhasha}, known by linguists as standard colloquial Bengali, is a written style that is comparatively easy and currently used in both formal and informal writings. 
Moreover, the writing form, syntax or choice of phrases differ greatly based on the origins: Bangladesh, West-Bengal (India) and Bengal province (British India). 
We create a corpus containing the writings of eight distinguished authors in Bengali literature.
A summary of the literature corpus is provided in Table~\ref{Tab-literature}.  
\textcolor{black}{The total number of unique words in this corpus is  \Tlide 30K. }

\begin{table*}[h]
\centering
\begin{tabular}{|l|l|l|l|l|l|}
\hline
\textbf{Author Name}                                                      & \textbf{Origin} & \textbf{Career} & \textbf{Dialect} & \textbf{Book \#}                                                                                                                                      & \textbf{Avg \# sent.}                                                        \\ \hline
\begin{tabular}[c]{@{}l@{}}Bankim \\ Chandra (BC)\end{tabular}            & Bengal          & 1866-1879       & Sadhu            & Novel: 14                                                                                                                                             & 3487                                                                            \\ \hline
\begin{tabular}[c]{@{}l@{}}Rabindranath\\  Tagore (RT)\end{tabular}       & Bengal          & 1883-1940       & Both             & \begin{tabular}[c]{@{}l@{}}Novel: 12\\ Short Story: 104\end{tabular}                                                                                  & \begin{tabular}[c]{@{}l@{}}4076\\ 196\end{tabular}                              \\ \hline
\begin{tabular}[c]{@{}l@{}}Sarat Chandra \\ Chatterjee (SC)\end{tabular}  & Bengal          & 1914-1931       & Sadhu            & \begin{tabular}[c]{@{}l@{}}Novel: 24\\ Short story: 29\end{tabular}                                                                                   & \begin{tabular}[c]{@{}l@{}}3495\\ 117\end{tabular}                              \\ \hline
\begin{tabular}[c]{@{}l@{}}Humayun \\ Ahmed (HA)\end{tabular}             & Bangladesh      & 1970-2011       & Cholito          & \begin{tabular}[c]{@{}l@{}}Novel: 103\\ Short story: 110\\ Long/Historical novel: 7\\ Himu: 23\\ Misir series:  20\\ Science fiction: 10\end{tabular} & \begin{tabular}[c]{@{}l@{}}4718\\ 462\\ 12451\\ 3489\\ 3062\\ 2740\end{tabular} \\ \hline
\begin{tabular}[c]{@{}l@{}}Sunil \\ Gangopadhyay\\  (SG)\end{tabular}     & West Bengal     & 1965-2012       & Cholito          & \begin{tabular}[c]{@{}l@{}}Novel: 8\\ Long/Historical novel: 6\\ Thriller series: 36\end{tabular}                                                     & \begin{tabular}[c]{@{}l@{}}6872\\ 17470\\ 3032\end{tabular}                     \\ \hline
\begin{tabular}[c]{@{}l@{}}Shirshendu\\ Mukhopadhyay\\  (SM)\end{tabular} & West Bengal     & 1967-           & Cholito          & \begin{tabular}[c]{@{}l@{}}Novel: 4\\ Thriller: 7\\ Long/Historical novel: 3\\ Teenage novel: 39\end{tabular}                                         & \begin{tabular}[c]{@{}l@{}}5897\\ 3904\\ 18185\\ 3858\end{tabular}              \\ \hline
\begin{tabular}[c]{@{}l@{}}Samaresh \\ Majumdar (SMM) \end{tabular}     & West Bengal     & 1976-           & Cholito          & \begin{tabular}[c]{@{}l@{}}Novel: 8\\ Short story: 72\end{tabular}                                                                                    & \begin{tabular}[c]{@{}l@{}}8704\\ 725\end{tabular}                              \\ \hline
\begin{tabular}[c]{@{}l@{}}Zafar Iqbal (ZI)\\ \end{tabular}                & Bangladesh      & 1982-           & Cholito          & \begin{tabular}[c]{@{}l@{}}Teenage novel: 11\\ Science fiction: 33\end{tabular}                                                                       & \begin{tabular}[c]{@{}l@{}}3587\\ 3283\end{tabular}                             \\ \hline
\end{tabular}
\caption{\label{Tab-literature} Distribution of authors and genres in Bengali literature corpus. }
\end{table*}

\paragraph{English literature fiction corpus:}

To show the effectiveness of \textit{Word2vec graph}, we perform various stylometry tasks on an English literature fiction dataset. It is a subset of the Project Gutenberg corpus, which is an extensive web catalog containing over fifty thousand e-books. 
All books have been manually cleaned to remove metadata, license information, and transcribers' notes. 
Since we do not have enough expertise in this domain and can not verify stylistic similarity among different writings from prior knowledge, we work with a small sample size in this dataset. 
Our version of the dataset consists of six categories of books from 54 authors. Overview of the dataset with book count is presented in Table \ref{tab-gutenberg}. \textcolor{black}{The total number of unique words in this corpus is  \Tlide 35K. }
\begin{table}
\centering

\begin{tabular}{|l|c|c|c|}
\hline
\textbf{Genre}               & 
\begin{tabular}[c]{@{}c@{}}\textbf{Book} \\ \textbf{\#}\end{tabular} & 
\begin{tabular}[c]{@{}c@{}}\textbf{Avg} \#\\ \textbf{sent.}\end{tabular} & \begin{tabular}[c]{@{}c@{}}\textbf{Avg} \#\\ \textbf{chapter}\end{tabular} \\ \hline
Detective \& mystery & 72    & 8695                                                      & 18.4                                                     \\
Love stories         & 64    & 17691                                                     & 25.1                                                     \\
Historical fiction   & 60    & 15496                                                     & 27.9                                                     \\
Adventure stories    & 71    & 9696                                                      & 24.5                                                     \\
Western stories      & 38    & 9955                                                      & 22.9                                                     \\
Science fiction      & 46    & 6803                                                      & 9.5                                                      \\ \hline
\end{tabular}
\caption{Overview of our Gutenberg corpus }
\label{tab-gutenberg}
\end{table}
\paragraph{Newspaper editorial corpus}
We also experiment with a Bengali newspaper editorial corpus to show comparative performance in a different domain.
We create the dataset from five prominent newspapers
in the Bengali language, such as Prothom Alo (Pr), Ittefaq (It), Jugantor (Ju), Inqilab (In), and Anandabazar (An). 
We only consider the editorial writings since they are supposed to be written by the editors/sub-editors of the newspaper and should contain their writing stylistics.
To reduce the topic and context dissimilarity among these articles, we crawl the data for a fixed period of time (May 2019 - November 2019)  for all newspapers. Table~\ref{Tab-newspaper} provides a summary of the corpus.
\textcolor{black}{Number of unique words in this corpus is  \Tlide 20K. }

\begin{table}[h]
\centering
\begin{tabular}{|l|l|l|c|}
\hline
\textbf{Newspaper} & \textbf{Origin} & \textbf{Dialect} & \textbf{\#} \\
\hline
Prothom Alo           & Bangladesh      & \textit{Cholito} & 254                    \\
Ittefaq                 & Bangladesh      & \textit{Sadhu}   & 254                    \\
Jugantor               & Bangladesh      & \textit{Cholito} & 222                    \\
Inqilab                 & Bangladesh      & \textit{Cholito} & 202                    \\
Anandabazar           & West Bengal     & \textit{Cholito} & 257     \\
\hline
\end{tabular}
\caption{\label{Tab-newspaper} Article counts and characteristics of different newspaper editorials}
\end{table}

%% file: sec_baseline.tex
\subsection{Baseline Methods}

Content-based features tend to be suitable for datasets with high topical diversity where datasets with a less topical variance, benefit more from style-based features~\cite{sari2018topic}. Therefore, we consider both Bag of words (unigram) feature set having Term Frequency Inverse Document Frequency (TF-IDF) score \& character n-grams to capture the content and stylometry feature set  (lexical, syntactic, sentiment, etc.) to find the stylistic signature of authors and utilize them as baseline methods.
\textcolor{black}{Also, we consider multilingual BERT representation of documents as a feature set. We employ the same classification and clustering techniques for all feature sets.}
\paragraph{Bag of word feature set with TF-IDF score: }
First, we generate a vocabulary of unique words for each corpus. 
Since the vocabulary size is large and many words have a very small frequency, 
we only consider unigrams to avoid overfitting and reduce feature dimension. Similar to \textit{Word2vec graph}, we have considered both \textit{with-stopwords} and  \textit{without-stopwords} version for TF-IDF feature set.
Given $|D|$ documents in corpus, we compute TF-IDF score for each word $w$  in a document $d$ as  $tf\_idf(w,d) = tf(w, d) \times log(|D|/(df + 1))$,
where term frequency is $tf(w,d) =$ count of $w$ in $d$ / number of words in $d$ and document frequency is $df(w) =$ occurrence of $w$ in documents. 
\textcolor{black}{We consider F as the number of total words in the corresponding corpus. }
\textcolor{black}{We experiment with the top 100 tokens from the TF-IDF feature set selected using the chi-square test to offer more fair competition to the \textit{Word2vec graph} feature set.}

\paragraph{Stylometry feature set:}
We mostly incorporate lexical, character-based features from each document since they are applicable to any language/domain.
Semantic analysis can be difficult even with
language processors, particularly on unrestricted
texts~\cite{neal2018surveying} like literature, and NLP in Bengali is still not sufficient to extract these features. 
However, We include some syntax features, parts of speech (pos-tag) information, and the sentiment/emotion property of words to make the feature set more effective. There are nearly ~250 features (details in \href{https://drive.google.com/file/d/1z8hw7JbTdKWliL6SdQ9j-QiUMyN965TQ/view?usp=sharing}{supplementary material}) in the stylometry feature set. 

\paragraph{Character n-grams:}
\textcolor{black}{As a baseline, we have evaluated character n-grams of up to four characters. To create this feature set equivalent to TF-IDF, we have considered the top $F=30K$ features based on information gain. }

\paragraph{Multilingual BERT}
\textcolor{black}{Multilingual BERT (m-BERT), a single language model pre-trained from monolingual corpora in 104 languages, provides satisfactory results in document classification in cross-lingual contexts. We utilize sentence transformers~\cite{wolf-etal-2020-transformers}  to represent a document as a set of features (768 features extracted from the last hidden layer) using the mBERT model. }

\paragraph{Classification: SVM} 
\textcolor{black}{
Support Vector Machine (SVM) is a powerful classification model for text categorization tasks~\cite{neal2018surveying}. 
We utilize linear kernel as it works well with high dimensional text feature sets and is faster to train~\cite{leopold2002text}. Each feature set except \textit{Word2vec graph+graph words} provide $M \times F$ feature matrix of the dataset, which we directly use for classification. For \textit{Word2vec graph+graph word} feature set, we replace graph words set (length W) with their index in the vocabulary (we consider separate vocabulary for each corpus) and perform the classification using $M \times (F+W)$ feature matrix. Since the sample size for individual experiments is not that large, we use a 70\%-30\% train-test split ratio to incorporate sufficient test samples in all experiments~\cite{xu2018splitting}.
}

\paragraph{Clustering: k-means Clustering} 
\textcolor{black}{
We perform k-means clustering with the number of clusters $k$  equals to distinct author/genre present in the dataset used for that experiment.
We also experiment with $k$ less than the number of classes to discover a natural separation among writings and identify which writings are similar irrespective of their author/genre.
For \textit{Word2vec graph+graph words} feature set, we perform k-means clustering on the numerical feature list (graph structure related) first. Then we generate a similarity matrix $M \times M$ based on Jaccard similarity coefficient~\cite{niwattanakul2013using} from each pair of samples in the dataset. Finally, we perform spectral clustering~\cite{von2007tutorial} with $k$ clusters. We utilize both clusterings results (M*2 representation) and implement an SVM classifier to identify the predicted class of the samples. 
}

%% file: sec6_conclusion.tex
\section{Conclusion}

\textcolor{black}{
 In this paper, we have proposed the \textit{Word2vec graph}, a novel feature set to represent a document for literary analysis. We extract various features from the graph structure and use relevant word information to perform author attribution and genre detection. We have performed an extensive evaluation of our proposed feature set using three separate datasets: a rich Bengali literature corpus, an English fictional writing dataset, and a Bengali newspaper editorial dataset. We compare the performance of \textit{Word2vec graph} approach and its variations with several baselines that include word unigram, stylometry, character n-gram and multi-lingual BERT feature sets. Our model performs the best in genre detection and achieves almost equal performance as the m-BERT \& TF-IDF feature set in the other task,  even though it uses significantly lower number selected attributes than other competitive approaches.
We believe that  \textit{Word2vec graph} could be an efficient approach to represent any text document in NLP. 
}

%% file: main.bbl
\begin{thebibliography}{45}
\expandafter\ifx\csname natexlab\endcsname\relax\def\natexlab#1{#1}\fi

\bibitem[{Abbasi and Chen(2008)}]{abbasi2008writeprints}
Ahmed Abbasi and Hsinchun Chen. 2008.
\newblock Writeprints: A stylometric approach to identity-level identification and similarity detection in cyberspace.
\newblock \emph{ACM Transactions on Information Systems (TOIS)}, 26(2):1--29.

\bibitem[{Alelyani et~al.(2018)Alelyani, Tang, and Liu}]{alelyani2018feature}
Salem Alelyani, Jiliang Tang, and Huan Liu. 2018.
\newblock Feature selection for clustering: A review.
\newblock In \emph{Data Clustering}, pages 29--60. Chapman and Hall/CRC.

\bibitem[{Amasyal{\i} and Diri(2006)}]{amasyali2006automatic}
M~Fatih Amasyal{\i} and Banu Diri. 2006.
\newblock Automatic turkish text categorization in terms of author, genre and gender.
\newblock In \emph{International Conference on Application of Natural Language to Information Systems}, pages 221--226. Springer.

\bibitem[{Ardanuy and Sporleder(2015)}]{ardanay2015clustering}
Mariona~Coll Ardanuy and Caroline Sporleder. 2015.
\newblock Clustering of novels represented as social networks.
\newblock In \emph{Linguistic Issues in Language Technology, Volume 12, 2015-Literature Lifts up Computational Linguistics}.

\bibitem[{Argamon and Levitan(2005)}]{argamon2005measuring}
Shlomo Argamon and Shlomo Levitan. 2005.
\newblock Measuring the usefulness of function words for authorship attribution.
\newblock In \emph{Proceedings of the 2005 ACH/ALLC Conference}, pages 4--7.

\bibitem[{Burns and Burns(2008)}]{burns2008anova}
Robert~P Burns and Richard Burns. 2008.
\newblock \emph{Business research methods and statistics using SPSS}.
\newblock Sage.

\bibitem[{Can and Patton(2004)}]{can2004change}
Fazli Can and Jon~M Patton. 2004.
\newblock Change of writing style with time.
\newblock \emph{Computers and the Humanities}, 38(1):61--82.

\bibitem[{Chowdhury et~al.(2018)Chowdhury, Imon, and Islam}]{chowdhury2018comparative}
Hemayet~Ahmed Chowdhury, Md~Azizul~Haque Imon, and Md~Saiful Islam. 2018.
\newblock A comparative analysis of word embedding representations in authorship attribution of bengali literature.
\newblock In \emph{2018 21st International Conference of Computer and Information Technology (ICCIT)}, pages 1--6. IEEE.

\bibitem[{Das and Mitra(2011)}]{das2011author}
Suprabhat Das and Pabitra Mitra. 2011.
\newblock Author identification in bengali literary works.
\newblock In \emph{International Conference on Pattern Recognition and Machine Intelligence}, pages 220--226. Springer.

\bibitem[{Diederich et~al.(2003)Diederich, Kindermann, Leopold, and Paass}]{diederich2003authorship}
Joachim Diederich, J{\"o}rg Kindermann, Edda Leopold, and Gerhard Paass. 2003.
\newblock Authorship attribution with support vector machines.
\newblock \emph{Applied intelligence}, 19(1-2):109--123.

\bibitem[{Feria et~al.(2018)Feria, Balbin, and Bautista}]{feria2018constructing}
Miguel Feria, Juan~Paolo Balbin, and Francis~Michael Bautista. 2018.
\newblock Constructing a word similarity graph from vector based word representation for named entity recognition.
\newblock \emph{arXiv preprint arXiv:1807.03012}.

\bibitem[{Ghaemi and Feizi-Derakhshi(2016)}]{ghaemi2016tree}
Manizheh Ghaemi and Mohammad-Reza Feizi-Derakhshi. 2016.
\newblock Feature selection using forest optimization algorithm.
\newblock \emph{Pattern Recognition}, 60:121--129.

\bibitem[{Guthrie(2008)}]{guthrie2008unsupervised}
David Guthrie. 2008.
\newblock \emph{Unsupervised detection of anomalous text}.
\newblock Ph.D. thesis, Citeseer.

\bibitem[{Halvani et~al.(2016)Halvani, Winter, and Pflug}]{halvani2016authorship}
Oren Halvani, Christian Winter, and Anika Pflug. 2016.
\newblock Authorship verification for different languages, genres and topics.
\newblock \emph{Digital Investigation}, 16:S33--S43.

\bibitem[{Houvardas and Stamatatos(2006)}]{houvardas2006n}
John Houvardas and Efstathios Stamatatos. 2006.
\newblock N-gram feature selection for authorship identification.
\newblock In \emph{International conference on artificial intelligence: Methodology, systems, and applications}, pages 77--86. Springer.

\bibitem[{Islam et~al.(2018)Islam, Kabir, Islam, and Tasnim}]{islam2018authorship}
Md~Ashikul Islam, Md~Minhazul Kabir, Md~Saiful Islam, and Ayesha Tasnim. 2018.
\newblock Authorship attribution on bengali literature using stylometric features and neural network.
\newblock In \emph{2018 4th International Conference on Electrical Engineering and Information \& Communication Technology (iCEEiCT)}, pages 360--363. IEEE.

\bibitem[{Islam(2003)}]{islam2003banglapedia}
Sirajul Islam. 2003.
\newblock \emph{Banglapedia: national encyclopedia of Bangladesh}, volume~3.
\newblock Asiatic society of Bangladesh.

\bibitem[{Jonker and Volgenant(1986)}]{jonker1986hungarian}
Roy Jonker and Ton Volgenant. 1986.
\newblock Improving the hungarian assignment algorithm.
\newblock \emph{Operations Research Letters}, 5(4):171--175.

\bibitem[{Kar et~al.(2019)Kar, Aguilar, and Solorio}]{kar2019multi}
Sudipta Kar, Gustavo Aguilar, and Thamar Solorio. 2019.
\newblock Multi-view characterization of stories from narratives and reviews using multi-label ranking.
\newblock \emph{arXiv preprint arXiv:1908.09083}.

\bibitem[{Ke{\v{s}}elj et~al.(2003)Ke{\v{s}}elj, Peng, Cercone, and Thomas}]{kevselj2003n}
Vlado Ke{\v{s}}elj, Fuchun Peng, Nick Cercone, and Calvin Thomas. 2003.
\newblock N-gram-based author profiles for authorship attribution.
\newblock In \emph{Proceedings of the conference pacific association for computational linguistics, PACLING}, volume~3, pages 255--264. sn.

\bibitem[{Kessler et~al.(1997)Kessler, Nunberg, and Sch{\"u}tze}]{kessler1997automatic}
Brett Kessler, Geoffrey Nunberg, and Hinrich Sch{\"u}tze. 1997.
\newblock Automatic detection of text genre.
\newblock \emph{arXiv preprint cmp-lg/9707002}.

\bibitem[{Kim(2014)}]{kim2014convolutional}
Yoon Kim. 2014.
\newblock Convolutional neural networks for sentence classification.
\newblock \emph{arXiv preprint arXiv:1408.5882}.

\bibitem[{Kipf and Welling(2016)}]{kipf2016semi}
Thomas~N Kipf and Max Welling. 2016.
\newblock Semi-supervised classification with graph convolutional networks.
\newblock \emph{arXiv preprint arXiv:1609.02907}.

\bibitem[{Kjell et~al.(1994)Kjell, Woods, and Frieder}]{kjell1994discrimination}
Bradley Kjell, W~Addison Woods, and Ophir Frieder. 1994.
\newblock Discrimination of authorship using visualization.
\newblock \emph{Information processing \& management}, 30(1):141--150.

\bibitem[{Koppel et~al.(2011)Koppel, Schler, and Argamon}]{koppel2011authorship}
Moshe Koppel, Jonathan Schler, and Shlomo Argamon. 2011.
\newblock Authorship attribution in the wild.
\newblock \emph{Language Resources and Evaluation}, 45(1):83--94.

\bibitem[{Labatut and Bost(2019)}]{labatut2019extraction_review}
Vincent Labatut and Xavier Bost. 2019.
\newblock Extraction and analysis of fictional character networks: A survey.
\newblock \emph{ACM Computing Surveys (CSUR)}, 52(5):1--40.

\bibitem[{Leopold and Kindermann(2002)}]{leopold2002text}
Edda Leopold and J{\"o}rg Kindermann. 2002.
\newblock Text categorization with support vector machines. how to represent texts in input space?
\newblock \emph{Machine Learning}, 46(1):423--444.

\bibitem[{Mihalcea and Tarau(2004)}]{mihalcea2004textrank}
Rada Mihalcea and Paul Tarau. 2004.
\newblock Textrank: Bringing order into text.
\newblock In \emph{Proceedings of the 2004 conference on empirical methods in natural language processing}, pages 404--411.

\bibitem[{Mikolov et~al.(2013)Mikolov, Chen, Corrado, and Dean}]{mikolov2013efficient}
Tomas Mikolov, Kai Chen, Greg Corrado, and Jeffrey Dean. 2013.
\newblock Efficient estimation of word representations in vector space.
\newblock \emph{arXiv preprint arXiv:1301.3781}.

\bibitem[{Narayanan et~al.(2017)Narayanan, Chandramohan, Venkatesan, Chen, Liu, and Jaiswal}]{narayanan2017graph2vec}
Annamalai Narayanan, Mahinthan Chandramohan, Rajasekar Venkatesan, Lihui Chen, Yang Liu, and Shantanu Jaiswal. 2017.
\newblock graph2vec: Learning distributed representations of graphs.
\newblock \emph{arXiv preprint arXiv:1707.05005}.

\bibitem[{Neal et~al.(2018)Neal, Sundararajan, Fatima, Yan, Xiang, and Woodard}]{neal2018surveying}
Tempestt Neal, Kalaivani Sundararajan, Aneez Fatima, Yiming Yan, Yingfei Xiang, and Damon Woodard. 2018.
\newblock Surveying stylometry techniques and applications.
\newblock \emph{ACM Computing Surveys (CSUR)}, 50(6):86.

\bibitem[{Niwattanakul et~al.(2013)Niwattanakul, Singthongchai, Naenudorn, and Wanapu}]{niwattanakul2013using}
Suphakit Niwattanakul, Jatsada Singthongchai, Ekkachai Naenudorn, and Supachanun Wanapu. 2013.
\newblock Using of jaccard coefficient for keywords similarity.
\newblock In \emph{Proceedings of the international multiconference of engineers and computer scientists}, volume~1, pages 380--384.

\bibitem[{Perozzi et~al.(2014)Perozzi, Al-Rfou, and Skiena}]{perozzi2014deepwalk}
Bryan Perozzi, Rami Al-Rfou, and Steven Skiena. 2014.
\newblock Deepwalk: Online learning of social representations.
\newblock In \emph{Proceedings of the 20th ACM SIGKDD international conference on Knowledge discovery and data mining}, pages 701--710.

\bibitem[{Phani et~al.(2015)Phani, Lahiri, and Biswas}]{phani2015authorship}
Shanta Phani, Shibamouli Lahiri, and Arindam Biswas. 2015.
\newblock Authorship attribution in bengali language.
\newblock In \emph{Proceedings of the 12th International Conference on Natural Language Processing}, pages 100--105.

\bibitem[{Reagan et~al.(2016)Reagan, Mitchell, Kiley, Danforth, and Dodds}]{reagan2016emotional}
Andrew~J Reagan, Lewis Mitchell, Dilan Kiley, Christopher~M Danforth, and Peter~Sheridan Dodds. 2016.
\newblock The emotional arcs of stories are dominated by six basic shapes.
\newblock \emph{EPJ Data Science}, 5(1):31.

\bibitem[{Sari et~al.(2018)Sari, Stevenson, and Vlachos}]{sari2018topic}
Yunita Sari, Mark Stevenson, and Andreas Vlachos. 2018.
\newblock Topic or style? exploring the most useful features for authorship attribution.
\newblock In \emph{Proceedings of the 27th International Conference on Computational Linguistics}, pages 343--353.

\bibitem[{Stamatatos(2009)}]{stamatatos2009survey}
Efstathios Stamatatos. 2009.
\newblock A survey of modern authorship attribution methods.
\newblock \emph{Journal of the American Society for information Science and Technology}, 60(3):538--556.

\bibitem[{Stamatatos et~al.(2000)Stamatatos, Fakotakis, and Kokkinakis}]{stamatatos2000automatic}
Efstathios Stamatatos, Nikos Fakotakis, and George Kokkinakis. 2000.
\newblock Automatic text categorization in terms of genre and author.
\newblock \emph{Computational linguistics}, 26(4):471--495.

\bibitem[{Stanchev(2014)}]{stanchev2014creating}
Lubomir Stanchev. 2014.
\newblock Creating a similarity graph from wordnet.
\newblock In \emph{Proceedings of the 4th International Conference on Web Intelligence, Mining and Semantics (WIMS14)}, pages 1--11.

\bibitem[{Von~Luxburg(2007)}]{von2007tutorial}
Ulrike Von~Luxburg. 2007.
\newblock A tutorial on spectral clustering.
\newblock \emph{Statistics and computing}, 17(4):395--416.

\bibitem[{Wolf et~al.(2020)Wolf, Debut, Sanh, Chaumond, Delangue, Moi, Cistac, Rault, Louf, Funtowicz, Davison, Shleifer, von Platen, Ma, Jernite, Plu, Xu, Scao, Gugger, Drame, Lhoest, and Rush}]{wolf-etal-2020-transformers}
Thomas Wolf, Lysandre Debut, Victor Sanh, Julien Chaumond, Clement Delangue, Anthony Moi, Pierric Cistac, Tim Rault, Rémi Louf, Morgan Funtowicz, Joe Davison, Sam Shleifer, Patrick von Platen, Clara Ma, Yacine Jernite, Julien Plu, Canwen Xu, Teven~Le Scao, Sylvain Gugger, Mariama Drame, Quentin Lhoest, and Alexander~M. Rush. 2020.
\newblock \href {https://www.aclweb.org/anthology/2020.emnlp-demos.6} {Transformers: State-of-the-art natural language processing}.
\newblock In \emph{Proceedings of the 2020 Conference on Empirical Methods in Natural Language Processing: System Demonstrations}, pages 38--45, Online. Association for Computational Linguistics.

\bibitem[{Worsham and Kalita(2018)}]{worsham2018genre}
Joseph Worsham and Jugal Kalita. 2018.
\newblock Genre identification and the compositional effect of genre in literature.
\newblock In \emph{Proceedings of the 27th International Conference on Computational Linguistics}, pages 1963--1973.

\bibitem[{Xu and Goodacre(2018)}]{xu2018splitting}
Yun Xu and Royston Goodacre. 2018.
\newblock On splitting training and validation set: a comparative study of cross-validation, bootstrap and systematic sampling for estimating the generalization performance of supervised learning.
\newblock \emph{Journal of Analysis and Testing}, 2(3):249--262.

\bibitem[{Zhao and Zobel(2007)}]{zhao2007searching}
Ying Zhao and Justin Zobel. 2007.
\newblock Searching with style: Authorship attribution in classic literature.
\newblock In \emph{Proceedings of the thirtieth Australasian conference on Computer science-Volume 62}, pages 59--68. Australian Computer Society, Inc.

\bibitem[{Zuo et~al.(2017)Zuo, Zhang, and Xia}]{zuo2017textrank-w2v}
Xiaolei Zuo, Silan Zhang, and Jingbo Xia. 2017.
\newblock The enhancement of textrank algorithm by using word2vec and its application on topic extraction.
\newblock In \emph{Journal of Physics: conference series}, volume 887, page 012028. IOP Publishing.

\end{thebibliography}
